# Spin-Orbit Interaction of Nuclear Shell Structure

Xiaobin Wang

*Seagate Technology,*

*7801 Computer Avenue South, Bloomington, Minnesota 55435, U.S.A.*

Zhengda Wang,

*Institute of Modern Physics, the Chinese Academy of Science,*

*Lanzhou, 730000, P. R. China*

Xiaochun Wang, Xiaodong Zhang

*The University of Texas, M. D. Anderson Cancer Center*

*1515 Holcombe Blvd, Houston, TX 77030, U.S.A.*

ABSTRACT

Single particle spin-orbit interaction energy problem in nuclear shell structure is solved through negative harmonic oscillator in the self-similar-structure shell model (SSM) [4] and considering quarks' contributions on single particle spin and orbit momentum. The paper demonstrates that single particle motion in normal nuclei is described better by SSM negative harmonic oscillator than conventional shell model positive harmonic oscillator[1][2][3]. The proposed theoretical formula for spin orbit interaction energy agrees well to experiment measurements.

The nuclear shell model is the foundation of nuclear structure theory. The problem of spin-orbit interaction of single particle motion in nuclear shell model has not been solved for long time. The general formula of spin-orbit interaction energy $\varepsilon_{nlj}^{s\cdot L}$ can be derived from the relativistic quantum mechanics and has been successfully used in describing electron motion in atomic shell structure. However, when applied to conventional nuclear shell model (SM), the general formula is not effective in describing nucleon single particle spin-orbit interaction. Two major problems exist: 1) spin orbit interaction energy in conventional shell model $\varepsilon_{nlj}^{s\cdot L}(SM)$ is positive for $j = l + 1/2$ and negative for $j = l - 1/2$. The sign of the theoretical formula is opposite to the sign of the experimental data $\varepsilon_{nlj}^{s\cdot L}(EXP)$ on normal nucleus. 2) The calculated spin-orbit interaction energy $\varepsilon_{nlj}^{s\cdot L}(SM)$ magnitude is much smaller (an order of magnitude) than that of the experimental value $\varepsilon_{nlj}^{s\cdot L}(EXP)$. In this paper the above two problems are solved by the proposed self-similar-structure shell model (SSM) and considering quarks contributions to spin and orbit momentum.

Single particle motion in self-similar shell model (SSM) is negative harmonic oscillator [4]. The sign of SSM spin-orbit energy $\varepsilon_{nlj}^{s\cdot L}(SSM)$ is opposite to that of SM spin orbit interaction energy $\varepsilon_{nlj}^{s\cdot L}(SM)$. It agrees to the sign of experiment measurement value $\varepsilon_{nlj}^{s\cdot L}(EXP)$. In addition, here we consider that a single nuclear particle consists of three quarks and three quarks participate in spin-orbit interactions equally. One single nuclear particle spin corresponds to three spins of the three quarks and one single nuclear particle orbit angular momentum corresponds to three orbit angular momentums of three quarks. Thus there is a factor of $3^2$ for the application of general spin-orbit interaction energy formula to nuclear single particle motion. This factor is important for theoretical prediction of the magnitude of SSM spin-orbit interaction value $\varepsilon_{nlj}^{s \cdot L}(SSM)$. Conventional nuclear shell model did not consider that a single particle (nucleon) consists of three quarks, thus the calculated spin orbit interaction energy $\varepsilon_{nlj}^{s \cdot L}(SM)$ is much smaller than the experiment value. Notice that self-similar-structure shell model negative harmonic circular oscillating frequency $\omega_{nl}^{SSM}$ depends upon the single particle configuration [4] so that the problem of spin-orbit interaction energy $\varepsilon_{nlj}^{s \cdot L}$ changing with orbit angular momentum L is also solved naturally.

The proposed formula for spin-orbit interaction energy is:

$$\varepsilon_{nlj}^{s \cdot L}(SSM) = 3^2 \left[ \frac{1}{2mc^2} \frac{1}{r} \frac{dU_{os}^{SSM}(nl,r)}{dr} \right] \vec{s} \cdot \vec{L} \qquad (1)$$

where $m$ is the single particle mass, $c$ is the speed of light, $U_{os}^{SSM}(nl,r)$ is the single particle potential. Exclusive of factor $3^2$, formula (1) can be derived from relativistic quantum mechanics. $3^2$ Factor is due to the summation of three quarks contributions to nucleon spin and orbit angular momentum, as discussed in previous paragraph. The single particle potential $U_{os}^{SSM}(nl,r)$ is the negative harmonic oscillator potential of self-similar-structure shell model [4]:

$$U_{os}^{SSM} = -\frac{m(\omega_{nl}^{SSM})^2 r^2}{2} \qquad (2)$$

where $\omega_{nl}^{SSM}$ is the single particle configuration dependent negative harmonic oscillating frequency. Substituting (2) to (1), the energy level for spin-orbit interaction is:

$$\varepsilon_{nlj}^{s \cdot L}(SSM) = -\frac{3^2(\omega_{nl}^{SSM}\hbar)^2 n_{ls}}{2mc^2\hbar^2}$$

$$n_{ls} = \frac{l}{2}\hbar^2 \quad \text{for} \quad j = l+1/2 \qquad (3)$$

$$n_{ls} = -\frac{(l+1)}{2}\hbar^2 \quad \text{for} \quad j = l-1/2$$

In the calculation of the total spin-orbit interaction $\sum_{1}^{A}\varepsilon_{nlj}^{s\cdot L}$ for normal nuclei (where A is the total number of nucleons), only $\varepsilon_{nlj}^{s\cdot L}$ of the outside single particle orbits are considered because $\varepsilon_{nlj}^{s\cdot L}$ of inside single particle orbits cancel each other. According to liquid drop model (DM) [5], the charge distribution is uniform in nucleus. The mean square root radius of outside single particle (neutron and proton) orbit is equal to the liquid drop radius.

$$R_{DM}^2 = \left[r_{r.m.s}^{nl}\right]^2 = \frac{(n+3/2)\hbar}{m\omega_{nl}} \qquad (4)$$

The spin-orbit interaction energy in SSM can be written as:

$$\varepsilon_{nlj}^{s\cdot L}(SSM) = -\frac{3^2}{2mc^2}\left[\frac{(n+3/2)\hbar^2}{mR_{DM}^2}\right]^2 \frac{n_{ls}}{\hbar^2} \qquad (5)$$

The DM radius is given as $R_{DM} = r_0 A^{1/3}$ with $r_0 = 1.2\,fm$. For liquid drop model, the binding energy $E_{DM}^0$ is a smooth function of nucleons number A and protons number Z. $\varepsilon_{nlj}^{s\cdot L}$ dependence upon A and Z is fluctuating. $\sum_{1}^{A}\varepsilon_{nlj}^{s\cdot L}$ correction is required for binding energy

$$E_{DM}(A,Z) = E_{DM}^0(A,Z) - \sum_{1}^{A}\varepsilon_{nlj}^{s\cdot L} \qquad (6)$$

Table 1 shows calculated SSM spin orbit interaction energy and its comparison to experimental measurement. Ten nuclei are selected. Most of them are magic number nuclei in order to investigate the effects of $\sum_{1}^{A}\varepsilon_{nlj}^{s\cdot L}$ on magic nuclei. In the calculation, spin orbit interaction energy (5) is:

$$\varepsilon_{nlj}^{s \cdot L}(SSM) = -2.395 \times 10^{-3} \left[ \frac{(n+3/2)41.444}{(1.2A^{1/3})^2} \right]^2 n_{s \cdot l} (MeV)$$

$$n_{s \cdot l} = \begin{cases} l & j = l+1/2 \\ -(l+1) & j = l-1/2 \end{cases}$$

(7)

Notice there is no fitting parameter in our calculation of SSM spin orbit interactions. The liquid drop energy is Weizacker formual [5]:

$$E_{DM}^0(A,Z) = a_v A - a_s A^{2/3} - a_c Z^2 A^{-1/3} - a_\alpha (A/2 - Z) A^{-1} + a_p \delta A^{-1/2}$$

$$a_v = 15.835 MeV, a_s = 18.33 MeV, a_c = 0.714 MeV, a_\alpha = 92.8 MeV, a_p = 11.2 MeV$$

$$\delta = \begin{cases} 1 & \text{even even nuclei} \\ 0 & \text{odd nuclei} \\ -1 & \text{odd odd nuclei} \end{cases}$$

(8)

$E_{exp}$ in table 1 is the experiment measured binding energy. It can be seen that SSM spin orbit interaction energy agrees to experiment measurement very well.

Long-standing problem of single particle motion spin-orbit interaction in nuclear shell model is solved. The solution is based upon negative harmonic oscillator in a self-similar-structure shell model and considers quarks' contributions on single particle spin and orbit momentum. The proposed spin orbit interaction energy formula agrees well to experiment measurements. The solution not only improves the theoretical understanding of spin-orbit interaction in nuclear shell model, but also demonstrates single particle motion in nuclear shell model is better described by negative harmonic oscillator. The solution also helps to improve the binding energy formula of liquid drop model.

**TABLE CAPTIONS**

TABLE. 1: Total spin-orbit interaction energy of SSM model, its correction to liquid drop model and its comparison to experiment measurement.

| nuclei(N,Z) | $_{82}Pb^{126}$ | $_{72}Hf^{112}$ | $_{54}Xe^{82}$ | $_{58}Ce^{72}$ | $_{38}Sr^{50}$ | $_{26}Fe^{28}$ | $_{20}Ca^{28}$ | $_{20}Ca^{16}$ | $_{8}O^{10}$ | $_{8}O^{6}$ |
|---|---|---|---|---|---|---|---|---|---|---|
| $\sum_{1}^{A}\varepsilon_{nlj}^{s \cdot L}(MeV)$: | $-11.71$ | $-0.81$ | $-8.92$ | $-1.27$ | $-6.55$ | $-8.28$ | $-5.53$ | $-2.45$ | $-3.58$ | $-1.47$ |
| $E_{DM}^{0}-E_{\exp}(MeV)$: | $-11.72$ | $-1.33$ | $-10.73$ | $-3.27$ | $-6.20$ | $-6.44$ | $-5.94$ | $-6.09$ | $-2.62$ | $-6.11$ |
| $E_{DM}-E_{\exp}(MeV)$: | $-0.01$ | $-0.52$ | $-1.81$ | $-2.00$ | $+0.35$ | $+1.84$ | $-0.42$ | $-3.64$ | $+0.96$ | $-4.64$ |

(Table 1. Wang et al for *PRL*)